\documentclass[sigconf]{acmart}

\setcopyright{none}  
\renewcommand\footnotetextcopyrightpermission[1]{}  

\acmConference[Augmented Educators and AI(CHI 2025 Workshop)]{Augmented Educators and AI: Shaping the Future of Human-AI Collaboration in Learning on CHI 2025 Workshop}{April 26,2025}{Yokohama, JAPAN}

\AtBeginDocument{%
  }

\begin{document}

\title[How Do Teachers Create Pedagogical Chatbots?: Current Practices and Challenges]{How Do Teachers Create Pedagogical Chatbots?:\\ Current Practices and Challenges}

\author{Minju Yoo}
\email{minjuu613@ewhain.net}
\orcid{0009-0000-2964-0464}
\affiliation{%
  \institution{Ewha Womans University}
  \city{Seoul}
  \country{Republic of Korea}
}

\author{Hyoungwook Jin}
\email{jinhw@kaist.ac.kr}
\orcid{0000-0003-0253-560X}
\affiliation{%
  \institution{School of Computing, KAIST}
  \city{Daejeon}
  \country{Republic of Korea}
}

\author{Juho Kim}
\email{juhokim@kaist.ac.kr}
\orcid{0000-0001-6348-4127}
\affiliation{%
  \institution{School of Computing, KAIST}
  \city{Daejeon}
  \country{Republic of Korea}
}


\begin{abstract}
AI chatbots have emerged as promising educational tools for personalized learning experiences, with advances in large language models (LLMs) enabling teachers to create and customize these chatbots for their specific classroom needs. However, there is a limited understanding of how teachers create pedagogical chatbots and integrate them into their lessons. Through semi-structured interviews with seven K-12 teachers, we examined their practices and challenges when designing, implementing, and deploying chatbots. Our findings revealed that teachers prioritize developing task-specific chatbots aligned with their lessons. Teachers engaged in various creation practices and had different challenges; novices in chatbot creation struggled mainly with initial design and technical implementation, while experienced teachers faced challenges with technical aspects and analyzing conversational data. Based on these insights, we explore approaches to supporting teachers' chatbot development and opportunities for designing future chatbot creation systems. This work provides foundational insights from teachers that can empower teacher-created chatbots, facilitating AI-augmented teaching.
\end{abstract}



\begin{teaserfigure}
   \includegraphics[width=\textwidth]{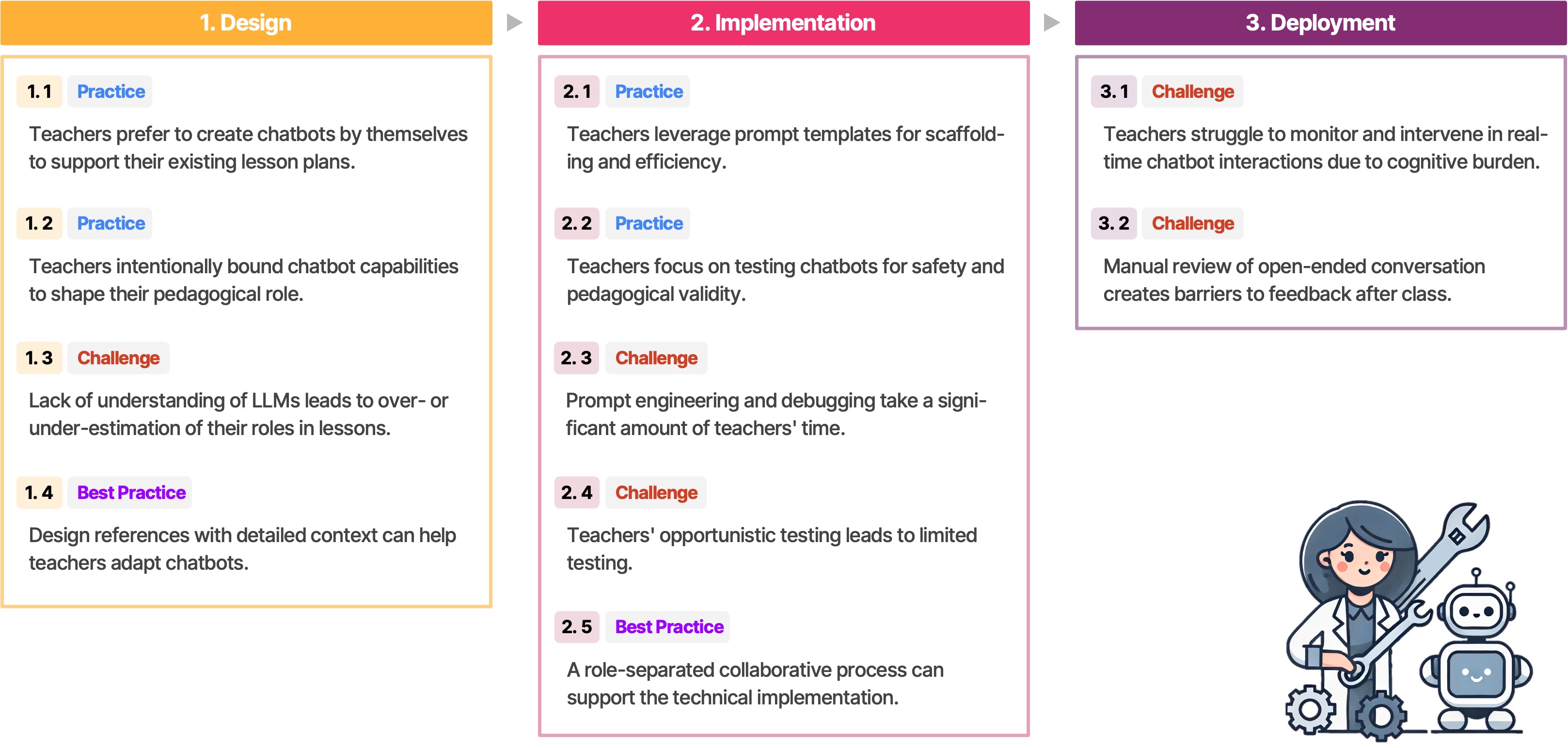}
   \caption{Our findings from semi-structured interviews with seven K-12 teachers highlight their challenges and practices throughout the chatbot creation process.}
   \Description{A figure of findings from semi-structured interviews with seven K-12 teachers, organized into three labeled boxes corresponding to different chatbot creation stages: design, implementation, and deployment. Within each stage, our findings are categorized into three groups: practices, challenges, and best practices.}
   \label{fig:teaser}
 \end{teaserfigure}

\maketitle
{\small
\noindent ©  This paper was adapted for the \textit{CHI 2025 Workshop on Augmented Educators and AI: Shaping the Future of Human and AI Cooperation in Learning},
held in Yokohama, Japan on April 26, 2025. This work is licensed under the Creative Commons Attribution 4.0 International License (CC BY 4.0).}

\newcommand{\category}[1]{[#1]}

\section{Introduction}
Many schools and institutes are exploring the use of AI chatbots in education to provide personalized, accessible, and scalable learning experiences. These efforts are boosted by large language models (LLMs) that can process and generate natural language. LLM's language processing capacity enabled the creation of highly interactive chatbots that can flexibly converse with students under diverse pedagogies, such as one-on-one tutoring~\cite{qi2023conversational}, learning by teaching~\cite{jin2024teach, schmucker2024ruffle}, and discussion-based learning~\cite{nguyen2024simulating, liu2024peergpt}. Beyond enhancing the quality of chatbots for these open-ended instructional methods, LLMs also empower teachers and educators to create and customize chatbots for their classrooms. The growing number of online workshops\footnote{https://gse-it.stanford.edu/project/build-bot-workshop-series} and commercial platforms\footnote{https://openai.com/index/introducing-gpts/} dedicated to building or using pre-designed educational chatbots reflects the increasing demand and interest in this technology.

However, little is known about how teachers at the forefront of education create and integrate LLM-based chatbots into their classrooms. Previous work has examined conceptual frameworks for designing pedagogical chatbots and teachers' roles in AI-supported learning environments~\cite{kim2024leading, lan2024teachers} and explored tools for teachers' creation~\cite{wang2024large, hedderich2024piece}. Now that LLM and chatbot creation are accessible to teachers, the next step is understanding how educators employ them over time and what challenges and workarounds emerge in real classroom settings. This work aims to take a glimpse of teachers' current practices and challenges in AI transformation and discuss how we can better support them.

We conducted semi-structured interviews with seven K-12 school teachers. The participants were either experienced teachers who had developed dozens of pedagogical chatbots by themselves or novice teachers who had not created chatbots before. For experienced teachers (n=3), we asked about their practical creation process and deployment experiences; for novice teachers (n=4), we asked them to use a commercial platform, Mizou~\footnote{https://mizou.com/education}, to build a chatbot of their interest and interviewed the hands-on experience. By interviewing these two groups, we aimed to observe novice teachers' initial barriers and expectations to chatbot creation and identfiy anecdotes, practices, and challenges of experienced teachers.

Our interview found that teachers prefer to create pedagogical chatbots tailored to specific class activities over a single versatile chatbot. Teachers wanted to develop chatbots that could fit into existing curricula and lesson plans instead of reorganizing curricula around a given chatbot so that they could keep their instructional strategies and practices in teaching. Teachers also wanted to limit what pedagogical chatbots can do and design specific roles for them that would support lessons while preventing students from overreliance on these tools. Lastly, teachers wanted to clearly divide their roles and chatbots' roles during classes and complement each other.

Teachers' practices and challenges resonate with their efforts to create tailored chatbots throughout the chatbot creation process. While novice teachers primarily struggled with initial design and technical implementation, expert teachers identified deeper insight and remaining challenges across the entire creation process and considered broader pedagogical implications of chatbot integration. Specifically, in the design stage, teachers carefully scope the roles and capabilities of chatbots, but their, especially novices’, limited understanding of LLMs often leads to over- and underestimating the capabilities of chatbots. During the implementation stage, both experienced and novice teachers engage in extensive prompt engineering and debugging to ensure that the chatbot can effectively support specific pedagogical actions and knowledge. When deploying chatbots into actual lessons, teachers desire to review students' interaction patterns in real-time and afterward to give students feedback, but open-ended interactions with chatbots make the review challenging. 

Based on the findings, we discuss how to enhance teachers' chatbot creation experience and propose future directions for system design. We believe that our insights from teachers' perspectives can lay the groundwork for the emerging field of LLM-based pedagogical chatbot design by teachers.

\section{Teacher Interview}
We conducted an hour-long semi-structured interview with teachers via Zoom to investigate how they create and use LLM-based pedagogical chatbots. The participants were compensated with KRW 20,000 (USD 14).

\subsection{Participants}
We recruited seven K-12 teachers through online teacher communities in Korea. Before the interviews, the participants indicated their background experience in creating LLM-based chatbots. We divided them into two groups (experienced and novice) based on their self-reported experience in chatbot creation in the interview application. Experienced teachers have experience building educational chatbots, while novice teachers have no experience but are interested in creating one. The following is an outline of the participants' experience.

\begin{itemize}
    \item \textbf{E1 (experienced):} He is an elementary school teacher with 15 years of teaching experience. He is knowledgeable about web programming and has developed two public websites. The first website is a pedagogical chatbot creation platform where teachers can write prompts to create chatbots and share them with their students for class usage. The platform provides six prompt samples and a prompt auto-generation feature to help novice teachers in prompt engineering. The second website is a gallery of chatbots and AI tools designed by teachers and students. He received 25 different ideas (e.g., a chatbot for studying multiplication tables and an image generator for a picture diary) from three public elementary schools and exhibited them on the website. He also regularly ran workshops to advertise his websites and help other teachers learn prompt engineering.

    \item \textbf{E2 (experienced):} He is an elementary school teacher with 17 years of teaching experience. He is knowledgeable about web programming and worked on a project led by a local office of education that aimed to build 101 pedagogical chatbots. The chatbots span 17 different subjects and have individual lesson plans and instructions crafted by school teachers. Some chatbots mimic a Korean historical character (history), teach complex arithmetic divisions (math), and advise daily exercise plans (physical education). He was the project's lead software engineer and worked closely with other teachers to create more than 30 chatbots from scratch. Since the project was large, there were teams dedicated to chatbot design, prompt engineering, and testing.

    \item \textbf{E3 (experienced):} She is an elementary school teacher with 7 years of teaching experience. She has experience in prompt engineering, chatbot testing, and deploying the chatbot in her classroom. She used a commercial platform to create a writing assistant that suggests ideas for arguments and supporting evidence based on students’ queries. She made the chatbot to help her students write essays within limited class times. She also conducted multiple teacher training courses on educational technologies, practical tips for using them, and the chatbot creation platforms.

    \item \textbf{N[1-4] (novice):} They were elementary and middle school teachers with varying teaching experience (1.3, 12, 15, and 24 years). They are familiar with ChatGPT and reported using it 2-3 times a week. Two teachers have experience using LLM in their classes. N2 designed a home economics class activity in which middle school students paired up and created a chatbot for daily clothing suggestions. N3 had a lightweight activity in which elementary school students conversed with an LLM-based chatbot for fun. All four teachers wanted to create chatbots, such as an English pronunciation corrector and a teaching assistant that students can ask questions in class.

\end{itemize}

\subsection{Interview Protocol}
Each interview began with an introduction to our research background, typical use cases of LLM-based chatbots in education, and interview procedures. We designed separate interview procedures for experienced and novice teachers to gain deeper insights from each group.

\subsubsection{Experienced Teachers}
We aimed to identify their long-standing design considerations, challenges, and workarounds for creating pedagogical chatbots. We first interviewed these experienced teachers in-depth about their experience and asked our research questions. Although the number of participants was small, their collaborative experience with numerous other teachers could provide generalizable insights.

\subsubsection{Novice Teachers}
We aimed to observe how novices approach chatbot creation and identify their expectations and misconceptions about LLM-based chatbots. For 40 minutes, each teacher used a commercial platform, Mizou, to design a pedagogical chatbot of their choice from scratch. The platform provided a basic template for teachers to fill in AI instructions, welcome messages, and rules, as well as a chat interface for testing. We intervened in their creation process when the teachers were stuck in prompt engineering or skipped testing chatbots. During and after the chatbot creation, we asked questions about their design rationales and thinking processes.

\section{Teachers' Practices and Challenges}
We identified teachers' challenges and practices for incorporating pedagogical chatbots into classes. We organized our findings by three heuristic stages of chatbot creation: design, implementation, and deployment. Each finding is categorized into three keywords: practices, challenges, and best practices.

\subsection{Design: Outlining Lesson Plans and Chatbot Specifications}
\subsubsection{\category{Practice} Teachers prefer to create chatbots by themselves to support their existing lesson plans.}
Regardless of their level of experience in chatbot creation, five teachers expressed the need to customize chatbots instead of using existing ones directly. Teachers wanted the agency to tailor chatbots to their existing lessons and teaching practices rather than redesign their lessons entirely to fit in chatbots. N4 explained, ``I would rather design a [writing assistant] chatbot that uses my strategies to help struggling students begin their essays rather than using generic AI writing tools.'' 

E2 also remarked, ``Chatbot design should start from identifying specific challenges, such as difficulties in time management during lessons or providing individual feedback. These challenges then clarify the design, roles, and behaviors of chatbots.'' For instance, E2 found that his students quickly lost interest in a book club because their book selection process was not fun. He created a chatbot to suggest book selection criteria and personalized book lists to make book exploration more engaging and have better discussions. E3 wanted to help students who spend too much time ideating and struggle to finish writing an essay within a class by creating a chatbot to guide students' essay brainstorming through questions.

\subsubsection{\category{Practice} Teachers intentionally bound chatbot capabilities to shape their pedagogical role.}
E1 stated ``A versatile chatbot across various subjects is not considered effective for both teaching and learning.'' E1 explained that limiting chatbot capabilities can prevent distraction and overreliance, helping students engage in lessons. Following this approach, N3 and N4 restricted their chatbots to subject-specific content (e.g., \textit{Do not respond to questions about subjects other than Social Sciences}). N4 stated, ``As students in my classroom tend to ask many off-topic questions, I wanted to prevent this issue rigorously.'' In addition, N2 sought to clearly define the chatbot's role as a complementary tool providing personalized answers to predefined questions. N2 explained, ``I needed a specific chatbot that could provide detailed solutions while I facilitated group discussions and managed overall classroom activities'' 

\subsubsection{\category{Challenge} Lack of understanding of LLMs leads to over- or under-estimation of their roles in lessons.}
While various examples of chatbots can inspire teachers' chatbot design, simply using chatbots may lack guidance in understanding the LLMs' capabilities, leading to barriers to adapting the examples to their actual context. E2 explained that novice teachers often overestimate chatbots' capabilities, assuming chatbots can address every challenging task. On the other hand, teachers might overlook potential chatbot design, although the technology could effectively support their needs. For instance, N4 was initially concerned about using chatbots in a writing class due to students copying AI-generated text, despite the need for chatbots to assist with personalized writing instruction. After understanding how prompts could limit AI-generated responses, N4 created a chatbot to guide questions on essay topics.

\subsubsection{\category{Best Practice} Design references with detailed context can help teachers adapt chatbots.}
E2 explained, ``Our project goal was to encourage teachers to identify potential chatbot applications within specific classroom contexts and adapt similar solutions for their own context.'' To achieve this goal, E2 suggested providing examples in lesson plan format, which is a familiar format for teachers. This format can provide detailed instructional context about when and how to integrate chatbots into lessons, including learning objectives, the purpose of chatbots, and teaching guidelines. Although this project is in its beta phase and requires further validation for real-world impact, it illustrates a promising approach to contextualized guidance for adapting chatbots to their unique classroom settings.  

\subsection{Implementation: Programming, Testing and Debugging Chatbots}

\subsubsection{\category{Practice} Teachers leverage prompt templates for scaffolding and efficiency.}
E1 remarked that prompt templates can lower the entry barrier for teachers who may hesitate about what to include. They serve as a frame of thinking, helping organize initially scattered requirements into a clear structure. E3 also explained, ``After a few trials, I could organize the rules into three parts: content knowledge, format, and teaching strategies. This categorization reduced my concern about missing rules I had not thought of.'' In addition, E2 highlighted the efficiency benefits, as stated, ``Based on my experience, if we could template various prompts, even regular teachers could create chatbots efficiently by just modifying specific keywords or topics.''

To structure prompt templates, E1 and E2 took different approaches. E1 used a fixed prompt template with general components---greeting messages, rules, conversation flow, and dialogue examples---to support teachers in creating diverse chatbots. E2, drawing from extensive experience creating various chatbots and prompts, believed that the structure of the prompt template should be tailored based on chatbots' roles or interaction types. For instance, in scenarios where students primarily ask questions and a chatbot responds (i.e., student-led Q\&A interactions), conversation flows become less crucial for the prompt, while answer formats are more important. However, for a chatbot designed to guide students through learning activities (i.e., chatbot-led instructional interactions), the templates required a detailed flow, including sequential questions and rules.

\subsubsection{\category{Practice} Teachers focus on testing chatbots for safety and pedagogical validity.}
After drafting initial prompts, teachers demonstrated common testing patterns while adapting their approaches based on their understanding of the subject matter and students. Their testing practices primarily focused on two aspects: safety testing for inappropriate messages, potential jailbreak trials, and validity testing to verify the pedagogical effectiveness of chatbot responses. For safety testing, teachers drew on their experiential knowledge to test inappropriate messages(E[1,3], N[3-4]). For validity testing, they checked the response format and content accuracy against their pedagogical intentions(E[2-3], N[1-4]). Teachers directly chatted with their chatbots to test responses in both single-turn and multi-turn scenarios. For instance, E3 structured her testing based on two student profiles she identified through experience: students with high and low motivation levels. While individual teachers had limited testing perspectives, E1 explained that some teachers address this limitation by peer-reviewing each other's chatbots.

\subsubsection{\category{Challenge} Prompt engineering and debugging take a significant amount of teachers' time.}
All teachers, even the experienced ones, spent substantial time crafting prompts that aligned with their pedagogical intentions. E1 noted, ``Multiple iterations were required to calibrate LLM responses from direct answers to appropriate hints for students.'' E1 and E3 observed that many teachers are unfamiliar with formulating effective prompts for LLMs and the iterative chatbot refinement process~\cite{lan2024teachers}. While prompt engineering can be an experiential process for teachers to discover LLM capabilities, it conflicts with their practical time constraints (E[1-3], N1, N[3-4]). As N4 noted, ``It would be challenging to continue this time-intensive process during the semester.'' It indicates the need for direct guidance on prompting engineering to reduce the time investment, such as feedback on LLM capabilities or suggestions for prompt refinement.

\subsubsection{\category{Challenge} Teachers' opportunistic testing leads to limited testing.}
All teachers conducted opportunistic testing rather than taking a systematic approach~\cite{zamfirescu2023why}. While this approach could help teachers address their immediate practical needs and contexts, we observed there would be several limitations in testing coverage. For instance, while chatbot conversations could be personalized for different students' interests or previous knowledge, these considerations were often overlooked. Teachers often struggled with determining educationally appropriate responses during testing. N3 stated, ``I am uncertain about the appropriate way for chatbots to handle students' use of profanity.'' This incomplete testing appeared to increase teachers' anxiety and potentially hindered future chatbot adoption (E[1-2], N[3-4]). E1 stated, ``Teachers sometimes felt pressure to guarantee the quality of their chatbot, worrying if chatbots generated inappropriate messages or students use the chatbot in a wrong way.''

\subsubsection{\category{Best Practice} A role-separated collaborative process can support the technical implementation.}
E2 introduced a collaborative design process that separates the roles of teachers and chatbot developers. In this workflow, teachers focus on pedagogical design - creating lesson plans and defining chatbot roles within these plans - while developers handle technical implementation. The collaboration continues as teachers identify necessary refinements through classroom testing and communicate chatbot specifications to developers. E2 explained, ``We documented both erroneous responses and specified ideal responses, rather than only reporting errors and their explanations.'' By analyzing the differences between the two responses, developers can better understand teachers' intentions and identify the desired implementation direction. In turn, developers can also provide feedback to teachers about technical aspects (i.e., prompt engineering and testing). E2 stated, ``As teachers often conducted testing without fully understanding the weaknesses of LLMs, we suggested feasible alternatives.'' While this approach leverages dedicated support teams for complex technical implementation, it suggests opportunities for scalable support systems~\cite{kim2024aineedsplanner}.

\subsection{Deployment: Conducting and Reviewing Lessons}

\subsubsection{\category{Challenge} Teachers struggle to monitor and intervene in real-time chatbot interactions due to cognitive burden.}
E1 emphasized the need for teachers to monitor real-time usage to handle corner cases or errors during the class. N3 also desired to monitor conversations to guide students in using the chatbot as intended for planned learning activities. While teachers' visits during student-AI tutor interactions relate to students' engagement~\cite{karumbaiah2023spatiotemporal}, managing multiple students' conversation histories and deciding when to intervene and how to act can create a cognitive burden. Current interfaces lack support for teachers to identify or address issues, requiring them to check individual conversation histories one by one. These limitations can hinder teachers' ability to provide timely interventions and ensure productive student-chatbot interactions.

\subsubsection{\category{Challenge} Manual review of open-ended conversation creates barriers to feedback after class.}
E1 noted, ``After the deployment, teachers should review whether chatbots and students engaged in conversation aligned with the learning objectives, identifying ways to improve the chatbot or the lesson.''  However, reviewing individual student-chatbot conversations requires significant time and effort. Due to the open-ended nature of conversational interactions, teachers need to manually click each student's profile and read all dialogues to assess the learning process (e.g., whether students are grasping new concepts or if they are getting distracted during activities). This could lead many teachers to either skip or minimize the feedback process. E1 observed that teachers' feedback on chatbot deployment often remained superficial, with teachers only mentioning general impressions like ``Students enjoyed the chatbot.'' The personal feedback from E3 is also more focused on the completion of the activity (``The activity went well.'') rather than identifying potential improvements or takeaways for future chatbot design.

\section{Discussion}
In this section, we discussed how to support teachers throughout the chatbot creation process and identified areas for future research.

\subsection{Understanding Chatbot Capabilities Through a Design Space}
We found that detailed references to lesson plans and chatbots have the potential to help teachers explore and adapt chatbots, addressing teachers' limited understanding of chatbot capabilities. To make these references more accessible for teachers' ideation process, we suggest developing a theoretical design space model for chatbot-supported teaching. While existing works focus on design frameworks in specific context (e.g., self-regulated learning~\cite{gazulla2022designing}), the design space can be defined across multiple domains by mapping relationships between pedagogical contexts (e.g., learning objectives, pedagogies) and chatbot interaction features (e.g., initiative role, output modality). One way to construct the dimensions would be to use an inductive method to collect various instances and components of chatbots and lessons through surveys for further characterization and distillation~\cite{zhang2024form}. By comprehensively organizing design dimensions, it could serve as a practical design guideline for chatbot creation. Simulating various design choices through dialogue examples could be another approach that can help teachers efficiently understand the chatbot capabilities and implications of different design dimensions.

\subsection{Interface Agent Support for Technical Implementation}
Inspired by E2's collaborative design process, we found that roles can be distinguished during chatbot creation: `teacher' who designs pedagogical requirements and `developer' who is responsible for technical implementation. It suggests an opportunity for system support with interface agents, as opposed to direct manipulation (i.e., prompt engineering)~\cite{shneiderman1997direct}. While agents can handle prompt engineering and debugging, future research should explore how teachers interact with agent outcomes considering the context of pedagogical chatbots. For example, agents could proactively verify teachers' input and provide feedback incorporating learning science theories or evaluate against predefined criteria through the LLM pipeline~\cite{kim2024evallm}. In addition, these agents would engage teachers by asking clarifying questions about ambiguous requirements. The system could further support comprehensive testing by recommending context-based test scenarios, allowing teachers to benefit from systematic testing perspectives with a lower workload while maintaining their agency in verification~\cite{jin2025teachtune}. This separation of roles would enable teachers to focus on their core expertise as pedagogical experts while maintaining their agency through verification processes.

\subsection{Modular Approach in Chatbot Creation Process}
We observed common patterns across chatbots designed by teachers, particularly in their educational approaches (e.g., avoiding direct answers) and safety concerns (e.g., sexual harassment, abuse). These interaction patterns indicate opportunities to modularize these interaction patterns, which can facilitate LLM-based chatbot development by reusing and sharing modules efficiently. For example, interaction modules could be constructed based on various pedagogies (e.g., Socratic questioning~\cite{alhossami2023socratic}), implemented through LLM pipelines~\cite{jin2024teach} or fine-tuned models~\cite{stowe2022controlled}. Teachers could build their interactions upon these modules instead of starting from scratch. More research is needed to explore how to define each interaction unit and how systems can effectively help teachers configure these modular components.

\subsection{Analyzing Student-chatbot Interactions for Classroom Orchestration}
We found that teachers need to analyze students' conversation data to monitor students' progress in real-time monitoring and to refine chatbots in subsequent classes. This finding extends previous research on teacher interactions in AI-supported classrooms, where monitoring AI tools helped augment teaching practices~\cite{an2020ta, karumbaiah2023spatiotemporal}, such as tracking students' conversation data and providing timely notifications for real-time interventions. While an existing framework has explored the balance between various aspects (e.g., autonomy and automation) in designing teaching augmentation systems~\cite{an2020ta}, further research could investigate how to provide appropriate features when teachers integrate chatbots into their classrooms. Beyond classroom support, post-class analysis tools could help teachers refine chatbot behaviors and overall lesson structure. For example, a dashboard that filters and analyzes conversations based on learning objectives could provide valuable information~\cite{kim2024designing}. Research is needed to explore how these analytical tools could support the continuous improvement of chatbots and generate actionable takeaways for future designs.

\bibliographystyle{ACM-Reference-Format}
\bibliography{references}

\end{document}